\documentclass[prb,aps,amsfonts,amssymb,amsmath,floats,floatfix,twocolumn,superscriptaddress,unsortedaddress]{revtex4}
\usepackage{graphics}
\usepackage{epsfig,graphicx,multirow}
\usepackage{graphicx}
\usepackage{amssymb}

\begin{document}

%\vspace{1cm} \noindent {\bf Submitted to Nature Physics (\today).}

\title{Two Gaps Make a High-Temperature Superconductor?}
\author{S. H\"{u}fner}
\affiliation{AMPEL, University of British Columbia, Vancouver, British Columbia, Canada V6T\,1Z4}
\affiliation{Department of Physics {\rm {\&}} Astronomy, University of British Columbia, Vancouver, British
Columbia, Canada V6T\,1Z1}
\author{M.A. Hossain}
\affiliation{AMPEL, University of British Columbia, Vancouver, British Columbia, Canada V6T\,1Z4}
\affiliation{Department of Physics {\rm {\&}} Astronomy, University of British Columbia, Vancouver, British
Columbia, Canada V6T\,1Z1}
\author{A. Damascelli}
\affiliation{AMPEL, University of British Columbia, Vancouver, British Columbia, Canada V6T\,1Z4}
\affiliation{Department of Physics {\rm {\&}} Astronomy, University of British Columbia, Vancouver, British
Columbia, Canada V6T\,1Z1}
\author{G.A. Sawatzky}
\affiliation{AMPEL, University of British Columbia, Vancouver, British Columbia, Canada V6T\,1Z4}
\affiliation{Department of Physics {\rm {\&}} Astronomy, University of British Columbia, Vancouver, British
Columbia, Canada V6T\,1Z1}

\begin{abstract}
One of the keys to the high-temperature superconductivity puzzle is the identification of the energy scales associated with the emergence of a
coherent condensate of superconducting electron pairs. These might provide a measure of the pairing strength and of the coherence of the
superfluid, and ultimately reveal the nature of the elusive pairing mechanism in the superconducting cuprates. To this end, a great deal of
effort has been devoted to investigating the connection between the superconducting transition temperature $T_c$ and the normal-state pseudogap
crossover temperature $T^*$. Here we present a review of a large body of experimental data that suggests a coexisting two-gap scenario, i.e.
superconducting gap and pseudogap, over the whole superconducting dome.
\end{abstract}

\date{\today}

\maketitle {\bf }

\begin{figure}[b!]
\centerline{\epsfig{figure=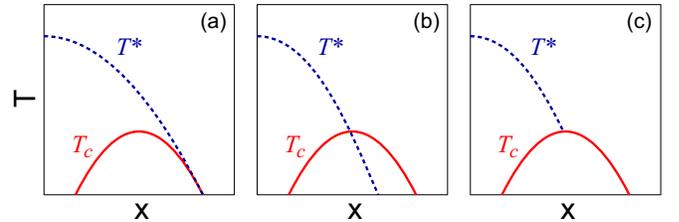,width=1\linewidth,clip=}} \vspace{-0.2cm} \caption{Various scenarios for the interplay of
pseudogap (blue dashed line) and superconductivity (red solid line) in the temperature-doping phase diagram of the HTSCs. While in (a) the
pseudogap merges gradually with the superconducting gap in the strongly overdoped region, in (b) and (c) the pseudogap lines intersects the
superconducting dome at about optimal doping (i.e., maximum $T_c$). In most descriptions, the pseudogap line is identified with a crossover with
a characteristic temperature $T^*$ rather than a phase transition; while at all dopings $T^*\!>\!T_c$ in (a), beyond optimal doping
$T^*\!<\!T_c$ in (b) and $T^*$ does not even exist in (c). Adapted from Ref.\onlinecite{8}.} \label{fig_1}
\end{figure}

Since their discovery,\cite{1} the copper-oxide high-$T_c$ superconductors (HTSCs) have become one of the most investigated class of
solids.\cite{1a,2,3,5,7,6c,6d,4,Campuzano,6b,8,6a,NatPhys,9,6,10,10a,6e,Schrieffer,10b,10c,7a,Uemura} However, despite the intense theoretical
and experimental scrutiny, an understanding of the mechanism that leads to superconductivity is still lacking. At the very basic level, what
distinguishes the cuprates from the conventional superconductors is the fact that they are doped materials, the highly atomic-like Cu 3$d$
orbitals give rise to strong electron correlations (e.g., the undoped parent compounds are antiferromagnetic Mott-Hubbard-like insulators), and
the superconducting elements are weakly-coupled two-dimensional layers (i.e., the celebrated square CuO$_2$ planes). Among the properties that
are unique to this class of superconducting materials, in addition to the unprecedented high superconducting $T_c$, the normal-state gap or {\it
pseudogap} is perhaps the most noteworthy. The pseudogap was first detected in the temperature dependence of the spin-lattice relaxation and
Knight shift in nuclear magnetic resonance and magnetic susceptibility studies.\cite{batlogg}  The Knight shift is proportional to the density
of states at the Fermi energy; a gradual depletion was observed below a cross-over temperature $T^*$, revealing the opening of the pseudogap
well above $T_c$\,on the underdoped side of the HTSC phase diagram (Fig.\,\ref{fig_1}). As the estimates based on thermodynamic quantities are
less direct than in spectroscopy we will, in the course of this review, mainly concentrate on spectroscopic results; more information on other
techniques can be found in the literature.\cite{5}

As established by a number of spectroscopic probes, among which primarily angle-resolved photoemission spectroscopy,\cite{loeserPG,dingPG} the
pseudogap manifests itself as a suppression of the  normal-state electronic density of states at $E_F$ exhibiting a momentum dependence
reminiscent of a $d_{x^2-y^2}$ functional form. For hole-doped cuprates, it is largest at Fermi momenta close to the antinodal region in the
Brillouin zone - i.e. around $(\pi,0)$ - and vanishes along the nodal direction - i.e. the $(0,0)$ to $(\pi,\pi)$ line. Note however that,
strictly speaking, photoemission and tunneling probe a suppression of spectral weight in the single-particle spectral function, rather than
directly of density of states; to address this distinction, which is fundamental in many-body systems and will not be further discussed here, it
would be very interesting to investigate the quantitative correspondence between nuclear magnetic resonance and single-particle spectroscopy
results. Also, no phase information is available for the pseudogap since, unlike the case of optimally and overdoped HTSCs,\cite{phaseSC} no
phase-sensitive experiments have been reported for the underdoped regime where $T^*\!\gg\!T_c$.
As for the doping dependence, the pseudogap $T^*$ is much larger than the superconducting $T_c$ in underdoped
samples, it smoothly decreases upon increasing the doping, and seems to merge with $T_c$ in the overdoped
regime, eventually disappearing together with superconductivity at doping levels larger than
$x\sim0.27$.\cite{4,Campuzano,5,6,6a,NatPhys,6b,6c,6d,6e,7,7a,8,9,10,10a,Schrieffer,10b,10c,Uemura}

In order to elaborate on the connection between pseudogap and high-$T_c$ superconductivity, or in other words
between the two energy scales $E_{pg}$ and $E_{sc}$ identified by $T^*$ and $T_c$ respectiely, let us start
by reminding that in conventional superconductors the onset of superconductivity is accompanied by the
opening of a gap at the chemical potential in the one-electron density of states.
\begin{figure*}[t]
\centerline{\epsfig{figure=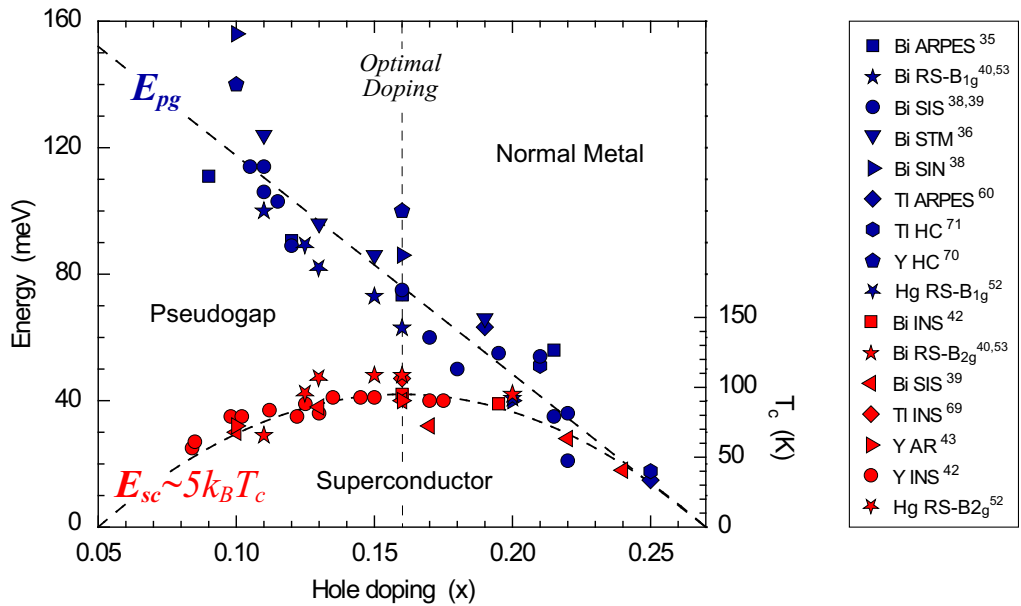,width=0.75 \linewidth,clip=}}  \vspace{-0.25cm} \caption{Pseudogap
($E_{pg}\!=\!2\Delta_{pg}$) and superconducting ($E_{sc}\!\sim\!5k_BT_c$) energy scales for a number of HTSCs with $T_c^{max}\!\sim\!95$\,K
(Bi2212, Y123, Tl2201, and Hg1201). The datapoints were obtained, as a function of hole doping $x$, by angle-resolved photoemission spectroscopy
(ARPES), tunneling (STM, SIN, SIS), Andreev reflection (AR), Raman scattering (RS), and heat conductivity (HC). On the same plot we are also
including the energy $\Omega_r$ of the magnetic resonance mode measured by inelastic neutron scattering (INS), which we identify with $E_{sc}$
because of the striking quantitative correspondence as a function of $T_c$. The data fall on two universal curves given by
$E_{pg}\!=\!E_{pg}^{max}(0.27 -x)/0.22$ and $E_{sc}\!=\!E_{sc}^{max}[1-82.6(0.16-x)^2]$, with
$E_{pg}^{max}\!=\!E_{pg}(x\!=\!0.05)\!=\!152\!\pm\!8$\,meV and $E_{sc}^{max}\!=\!E_{sc}(x\!=\!0.16)\!=\!42\!\pm\!2$\,meV (the statistical errors
refer to the fit of the selected datapoints; however, the spread of all available data would be more appropriately described by $\pm\!20$ and
$\pm\!10$\,meV, respectively).} \label{fig_2}
\end{figure*}
According to the Bardeen-Cooper-Schrieffer (BCS) theory of superconductivity,\cite{BCS} the gap energy
provides a direct measure of the binding energy of the two electrons forming a Cooper pair (the two-particle
bosonic entity that characterizes the superconducting state). It therefore came as a great surprise that a
gap, i.e. the pseudogap, was observed in the HTSCs not only in the superconducting state as expected from
BCS, but also well above $T_c$. Because of these properties and the hope it might reveal the mechanism for
high-temperature superconductivity, the pseudogap phenomenon has been very intensely investigated. However,
no general consensus has been reached yet on its origin, its role in the onset of superconductivity itself,
and not even on its evolution across the HTSC phase diagram.

As discussed in three recent papers on the subject,\cite{8,9,10} and here summarized in Fig.\,\ref{fig_1},
three different phase diagrams are usually considered with respect to the pseudogap line. While Millis
\cite{9} opts for a diagram like in Fig.\,\ref{fig_1}a, Cho \cite{10} prefers a situation where the pseudogap
line meets the superconducting dome at $x\simeq0.16$ (Fig.\,\ref{fig_1}b,c); Norman et al.\cite{8} provide a
comprehensive discussion of the three different possibilities. One can summarize some of the key questions
surrounding the pseudogap phenomenon and its relevance to high-temperature superconductivity as
follows:\cite{8,9,10}

\begin{description}
%\vspace{-0.2cm}
\item[ ]1. Which is the correct phase diagram with respect to the pseudogap line?
\vspace{-0.2cm}
\item[ ]2. Does the pseudogap connect to the insulator quasiparticle spectrum?
\vspace{-0.2cm}
\item[ ]3. Is the pseudogap the result of some one-particle band structure effect?
\vspace{-0.2cm}
\item[ ]4. Or, alternatively, is it a signature of a two-particle pairing interaction?
\vspace{-0.2cm}
\item[ ]5. Is there a true order parameter defining the existence of a pseudogap phase?
\vspace{-0.2cm}
\item[ ]6. Do the pseudogap and a separate superconducting gap coexist below $T_c$?
\vspace{-0.2cm}
\item[ ]7. Is the pseudogap a necessary ingredient for high-$T_c$  superconductivity?
%\vspace{-0.2cm}
\end{description}

\noindent In this communication we will revisit some of these questions, with specific emphasis on the one-
vs. two-gap debate. Recently, this latter aspect of the HTSCs has been discussed in great detail by Goss
Levi,\cite{levi} in particular based on scanning-tunneling microscopy data from various
groups.\cite{gomes,boyer,kohsaka} Here we will expand this discussion to include the plethora of experimental
results available from a wide variety of techniques. We will show that one fundamental and robust conclusion
can be drawn: the HTSC phase diagram is dominated by two energy scales, the superconducting transition
temperature $T_c$ and the pseudogap crossover temperature $T^*$, which converge to the very same critical
point at the end of the superconducting dome. Establishing whether this phenomenology can be conclusively
described in terms of a coexisting two-gap scenario, and what the precise nature of the gaps would be, will
require a more definite understanding of the quantities measured by the various probes.

%\vspace{-0.3cm}
\section{Emerging Phenomenology}
%\vspace{-0.3cm}

The literature on the HTSC superconducting gap and/or pseudogap is very extensive and still growing. In this situation it seems interesting to
go over the largest amount of data obtained from as many experimental techniques as possible, and look for any possible systematic behavior that
could be identified. This is the primary goal of this focused review.
\begin{table}[t]
\begin{tabular}
{|l|c|c|c|} \hline
\multicolumn{4}{|c|}{Optimally doped Bi2212 ($T_c\!\sim\!90$-$95$\,K)} \\ \hline  Experiment  & \multirow{1}{*}{Energy} & meV & Ref. \\
\hline
 ARPES - ($\pi,\!0$) peak & \multirow {7}{*}{$E_{pg}$} & 80 & \onlinecite{30b,24}\\
 Tunneling - STM && 70 &\onlinecite{10a,16}\\
 Tunneling - SIN && 85 & \onlinecite{15}\\
 Tunneling - SIS && 75 & \onlinecite{12,13}\\
 Raman - $B_{1g}$ && 65 & \onlinecite{22}\\
 Electrodynamics && 80 & \onlinecite{5,30g}\\
 \hline
Neutron  - ($\pi,\pi$) $\Omega_r$ & \multirow{5}{*}{$E_{sc}$} & 40 & \onlinecite{31}\\
Raman - $B_{2g}$ &&45 & \onlinecite{22}\\
Andreev &&45 & \onlinecite{35}\\
SIS - dip &&40 & \onlinecite{13}\\
\hline
\end{tabular}
\caption{\label{tab:table2}Pseudogap $E_{pg}$ and superconducting $E_{sc}$ energy scales (2$\Delta$) as
inferred, for optimally doped Bi2212, from different techniques and experiments. Abbreviations are given in
the main text, while the original references are listed.} \label{tab_1}
\end{table}
We want to emphasize right from the start that we are not aiming at providing exact quantitative estimates of
superconducting and pseudogap energy scales for any specific compound or any given doping. Rather, we want to
identify the general phenomenological picture emerging from the whole body of available experimental
data.\cite{4,5,6,6a,10a,11,11a,12,13,14,15,16,17,18,19,19a,20,21,22,23,24,25,26,27,28,29,29a,30,30a,30ab,30aa,30b,30c,30d,30e,30f,30g,31,32,32a,33,34,35,35a}

In the following, we will consider some of the most direct probes of low-energy, electronic excitations and
spectral gaps, such as angle-resolved photoemission (ARPES), scanning tunneling microscopy (STM),
superconductor/insulator/normal-metal (SIN) and superconductor/insulator/superconductor (SIS) tunneling,
Andreev reflection tunneling (AR), and Raman scattering (RS), as well as less conventional probes such as
heat conductivity (HC) and inelastic neutron scattering (INS). The emphasis in this review will be on
spectroscopic data because of their more direct interpretative significance; however, these will be checked
against thermodynamic/transport data whenever possible. With respect to the spectroscopic data, it will be
important to differentiate between single-particle probes such as ARPES and STM, which directly measure the
one-electron excitation energy $\Delta$ with respect to the chemical potential (on both side of $E_F$ in
STM), and two-particle probes such as Raman and inelastic neutron scattering, which instead provide
information on the particle-hole excitation energy 2$\Delta$. Note that the values reported here are those
for the `full gap' $2\Delta$ (associated with either $E_{sc}$ or $E_{pg}$), while frequently only half the
gap $\Delta$ is given for instance in the ARPES literature. In doing so one implicitly assumes that the
chemical potential lies half-way between the lowest-energy single-electron removal and addition states; this
might not necessarily be correct but appears to be supported by the direct comparison between ARPES and
STM/Raman results. A more detailed discussion of the quantities measured by the different experiments and
their interpretation will be provided in the following subsections. Here we would like to point out that
studies of $B_{2g}$ and $B_{1g}$ Raman intensity,\cite{6e,21,22} heat conductivity of nodal
quasiparticles,\cite{33,34} and neutron magnetic resonance energy $\Omega_r$,\cite{31} do show remarkable
agreement with superconducting or pseudogap energy scales as inferred by single-particle probes, or with the
doping dependence of $T_c$ itself. Thus they provide, in our opinion, an additional estimate of $E_{sc}$ and
$E_{pg}$.

As for the choice of the specific compounds to include in our analysis, we decided to focus on those HTSCs
exhibiting a similar value of the maximum superconducting transition temperature $T_c^{max}$, as achieved at
optimal doping, so that the data could be quantitatively compared without any rescaling. We have therefore
selected Bi$_2$Sr$_2$CaCu$_2$O$_{8+\delta}$ (Bi2212), YBa$_2$Cu$_3$O$_{7-\delta}$ (Y123),
Tl$_2$Ba$_2$CuO$_{6+\delta}$ (Tl2201), and HgBa$_2$CuO$_{4+\delta}$ (Hg1201), which have been extensively
investigated and are all characterized by $T_c^{max}\!\sim\!95$\,K.\cite{hiroshi}
\begin{figure}[b!]
\centerline{\epsfig{figure=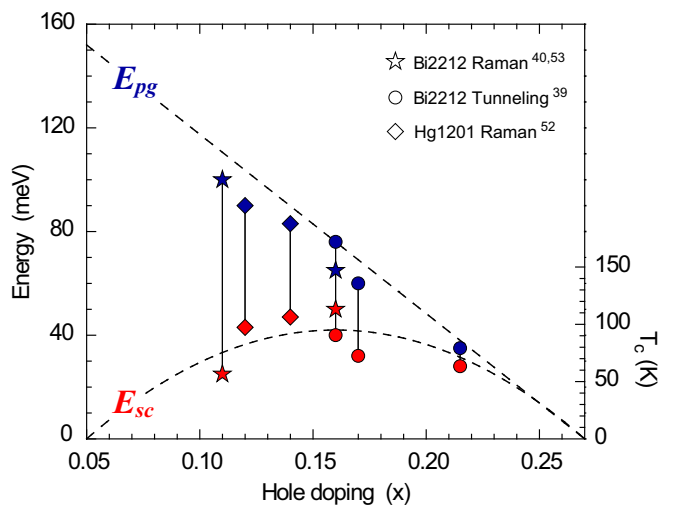,width=1\linewidth,clip=}} \vspace{-0.2cm}
\caption{Pseudogap $E_{pg}$ and superconducting $E_{sc}$ energy scales (2$\Delta$) as estimated, by a number
of probes and for different compounds, in one single experiment on the very same sample. These data provide
direct evidence of for the simultaneous presence of two energy scales, possibly two spectral gaps, coexisting
in the superconducting state. The superconducting and pseudogap lines are defined as in Fig.\,\ref{fig_2}.}
\label{fig_3}
\end{figure}
It should also be noted that while Bi2212 and Y123 are `bilayer' systems, i.e. their crystal structure
contains as a key structural element sets of two adjacent CuO$_2$ layers, Tl2201 and Hg1201 are structurally
simpler single CuO$_2$-layer materials. Therefore, this choice of compounds ensures that our conclusions are
generic to all HTSCs with a similar $T_c$, independent of the number of CuO$_2$ layers.

A compilation of experimental results for the magnitude of pseudogap ($E_{pg}\!=\!2\Delta_{pg}$) and
superconducting ($E_{sc}\!\sim\!5k_BT_c$) energy scales, as a function of carrier doping $x$, is presented in
Fig.\,\ref{fig_2} (only some representative datapoints are shown, such as not to overload the figure; similar
compilations were obtained also by a number of other authors).\cite{4,5,6,6a,21,28,30,31,33,35,pana,leni} The
data for these HTSCs with comparable $T_c^{max}\!\sim\!95$\,K fall on two universal curves: a straight line
for the pseudogap energy $E_{pg}\!=\!2\Delta_{pg}$, and a parabola for the superconducting energy scale
$E_{sc}\!\sim\!5k_BT_c$. The two curves converge to the same $x\sim0.27$ critical point at the end of the
superconducting dome, similar to the cartoon of Fig.\,\ref{fig_1}a. In order to summarize the situation with
respect to quantitative estimates of $E_{pg}$ and $E_{sc}$, we have listed in Table\,\ref{tab_1} the values
as determined by the different experimental techniques on optimally doped Bi2212 (with $T_c$ ranging from 90
to 95\,K). While one obtains from this compilation the average values of $E_{pg}\!\simeq\!76$\,meV and
$E_{sc}\!\simeq\!41$\,meV at optimal doping, the numbers do scatter considerably. Note also that these
numbers differ slightly from those given in relation to the parabolic and straight lines in Fig.\,\ref{fig_2}
(e.g., $E_{sc}^{max}\!=\!42$\,meV) because the latter were inferred from a fitting of superconducting and
pseudogap data over the whole doping range, while those in Table\,\ref{tab_1} were deduced from results for
optimally doped Bi2212 only. It is also possible to plot the pseudogap $E_{pg}$ and superconducting $E_{sc}$
energy scales as estimated simultaneously in one single experiment on the very same sample. This is done in
Fig.\,\ref{fig_3} for Raman, tunneling, and ARPES results from Bi2212 and Hg1201, which provide evidence for
the presence of two energy scales, or possibly two spectral gaps as we will discuss in greater detail below,
coexisting over the whole superconducting dome.

%\vspace{-0.3cm}
\subsection{Angle-resolved photoemission}
%\vspace{-0.3cm}

The most extensive investigation of excitation gaps in HTSCs has arguably been done by
ARPES.\cite{4,Campuzano,loeserPG,dingPG,24,25,26,27,28,kondo,ws,norman,kanigel,kmshen,29,29a,30,30a,30ab,30aa,30b,30c,30d,30e} This technique
provides direct access to the one-electron removal spectrum of the many-body system; it allows, for instance in the case of a BCS
superconductor,\cite{BCS} to measure the momentum dependence of the absolute value of the pairing amplitude $2\Delta$ via the excitation gap
$\Delta$ observed for single-electron removal energies, again assuming $E_F$ to be located half-way in the gap.\cite{4,Campuzano} This is the
same in some tunneling experiments such as STM, which however do not provide direct momentum resolution but measure on both side of
$E_F$.\cite{10a} The gap magnitude is usually inferred from the ARPES spectra from along the normal-state Fermi surface in the antinodal region,
where the $d$-wave gap is largest; it is estimated from the shift to high-binding energy of the quasiparticle spectral weight relative to the
Fermi energy. With this approach only one gap is observed below a temperature scale that smoothly evolves from the so-called pseudogap
temperature $T^*$ in the underdoped regime, to the superconducting $T_c$ on the overdoped side. We identify this gap with the pseudogap energy
scale $E_{pg}\!=\!2\Delta_{pg}$. This is also in agreement with recent investigations of the near-nodal ARPES spectra from single and double
layer Bi-cuprates,\cite{28,kondo,ws} which further previous studies of the underdoped cuprates' Fermi arc
phenomenology.\cite{norman,kanigel,kmshen} From the detailed momentum dependence of the excitation gap along the Fermi surface contour, and the
different temperature trends observed in nodal and antinodal regions, these studies suggest the coexistence of two distinct spectral gap
components over the whole superconducting dome: superconducting gap and pseudogap, dominating the response in the nodal and antinodal regions
respectively, which would eventually collapse to one single energy scale in the very overdoped regime.

%\vspace{-0.3cm}
\subsection{Tunneling}
%\vspace{-0.3cm}

The HTSCs have been investigated by a wide variety of tunneling
techniques,\cite{6a,10a,11,11a,12,13,14,15,16,17,18,19,19a,20} such as SIN,\cite{12,20} SIS,\cite{12,13,15},
STM,\cite{10a,14,16}, intrinsic tunneling,\cite{17,18,19,19a} and Andreev reflection, which is also a
tunneling experiment but involves two-particle rather than single-particle tunneling (in principle, very much
like SIS).\cite{6a,35,35a} All of these techniques, exception made for intrinsic tunneling,\cite{intrinsic}
are here represented either in the figures or table.

Similar to what was discussed for ARPES at the antinodes, there are many STM studies that report a pseudogap $E_{pg}$ smoothly evolving into
$E_{sc}$ upon overdoping.\cite{10a,gomes} In addition, a very recent temperature-dependent study of overdoped single-layer Bi-cuprate detected
two coexisting, yet clearly distinct, energy scales in a single STM experiment.\cite{boyer} In particular, while the pseudogap was clearly
discernible in the differential conductance exhibiting the usual large spatial modulation, the evidence for a spatially-uniform superconducting
gap was obtained by normalizing the low-temperature spectra by those just above $T_c\!\simeq\!15$\,K. These values have not been included in
Fig.\,\ref{fig_2} and \ref{fig_3} because $T_c\!\ll\!95$K; however, this study arguably provides the most direct evidence for the coexistence of
two distinct excitation gaps in the HTSCs.

One can regard Andreev reflection (pair creation in addition to a hole) as the inverse of a two-particle
scattering experiment such as Raman or INS. A different view is also possible: SIN tunneling goes over to AR
if the insulator layer gets thinner and thinner;\cite{6a} thus a SIN tunneling, as also STM, should give the
same result as AR. However while SIN and STM measure the pseudogap, AR appears to be sensitive to the
superconducting energy scale $E_{sc}$ (Fig.\,\ref{fig_2}). We can only conjecture that this has to do with
the tunneling mechanisms being actually different.

SIS tunneling experiments \cite{13} find $E_{pg}/E_{sc}\!\gtrsim\!1$ for Bi2212 at all doping levels. There
are however some open questions concerning the interpretation of the SIS experiments. This technique, which
exploits Josephson tunneling, measures pair spectra; the magnitude of $E_{pg}$ can readily be obtained from
the most pronounced features in the spectra.\cite{13} The signal related to $E_{sc}$ is seen as a `sideband'
on the $E_{pg}$ features; it seems not obvious why, if the $E_{sc}$ signal did originate from a state of
paired electrons, it would not show up more explicitly.

%\vspace{-0.3cm}
\subsection{Raman scattering}
%\vspace{-0.3cm}

Light scattering measures a two-particle excitation spectrum providing direct insights on the total energy needed to break up a two-particle
bound state or remove a pair from a condensate. Raman experiments can probe both superconducting and pseudogap energy scales, if one interprets
the polarization dependent scattering intensity in terms of different momentum averages of the $d$-wave-like gap functions: one peaked at
($\pi$,0) in $B_{1g}$ geometry, and thus more sensitive to the larger $E_{pg}$ which dominates this region of momentum space; the other at
($\pi$/2,$\pi$/2) in $B_{2g}$ geometry, and provides an estimate of the slope of the gap function about the nodes,
$\frac{1}{\hslash}\frac{d\Delta}{d{\bf k}}\!\!\mid_{n}$, which is more sensitive to the arguably steeper functional dependence of $E_{sc}$ out
of the nodes.\cite {6e, 21, 22, 23} One should note however that the signal is often riding on a high background, which might result in a
considerable error and data scattering. At a more fundamental level, while the experiments in the antinodal geometry allow a straightforward
determination of the gap magnitude $E_{pg}$, the nodal results need a numerical analysis involving a normalization of the Raman response
function over the whole Brillouin zone, a procedure based on a low-energy $B_{2g}$ sum rule (although also the $B_{2g}$ peak position leads to
similar conclusions).\cite {21} This is because a $B_{2g}$ Raman experiment is somewhat sensitive also to the gap in the antinodal direction,
where it picks up in particular the contribution from the larger pseudogap.

%\vspace{-0.3cm}
\subsection{Inelastic neutron scattering}
%\vspace{-0.3cm}

Inelastic neutron scattering experiments have detected the so-called $q\!=\!(\pi$,$\pi$) resonant magnetic
mode in all of the $T_c\!\simeq\!95$\,K HTSCs considered here.\cite{6} This resonance is proposed by some to
be a truly collective magnetic mode that, much in the same way as phonons mediate superconductivity in the
conventional BCS superconductors, might constitute the bosonic excitation mediating superconductivity in the
HTSCs. The total measured intensity, however, amounts to only a small portion of what is expected based on
the sum rule for the magnetic scattering from a spin $1/2$ system;\cite{6,6d,Uemura,31,32,32a}  this weakness
of the magnetic response should be part of the considerations in the modelling of magnetic-resonance mediated
high-$T_c$ superconductivity. Alternatively, its detection below $T_c$ might be a mere consequence of the
onset of superconductivity and of the corresponding suppression of quasiparticle scattering. Independently of
the precise interpretation, the INS data reproduced in Fig.\,\ref{fig_2} show that the magnetic resonance
energy $\Omega_r$ tracks very closely, over the whole superconducting dome, the superconducting energy scale
$E_{sc}\!\sim\!5 k_B T_c$ [similar behavior is observed, in the underdoped regime, also for the spin-gap at
the incommensurate momentum transfer ($\pi$,$\pi\pm\delta$)]\cite{joel}. Remarkable is also the
correspondence between the energy of the magnetic resonance and that of the $B_{2g}$ Raman peak. {\it Note
that while the $q\!=\!(\pi$,$\pi$) momentum transfer observed for the magnetic resonance in INS is a key
ingredient of most proposed HTSC descriptions, Raman scattering is a $q\!=\!0$ probe}. It seems that
understanding the connection between Raman and INS might reveal very important clues.

%\vspace{-0.3cm}
\subsection{Heat conductivity}
%\vspace{-0.3cm}

Heat conductivity data from Y123 and Tl2201 fall on to the pseudogap line. This is a somewhat puzzling result
because they have been measured at very low temperatures, well into the superconducting state, and should in
principle provide a measure of both gaps together if these were indeed coexisting below $T_c$. However,
similar to the $B_{2g}$ Raman scattering, these experiments are only sensitive to the slope of the gap
function along the Fermi surface  at the nodes, $\frac{1}{\hslash}\frac{d\Delta}{d{\bf k}}\!\!\mid_{n}$; the
gap itself is determined through an extrapolation procedure in which only one gap was assumed. The fact that
the gap values, especially for Y123, come out on the high side of the pseudogap line may be an indication
that an analysis with two coexisting gaps might be more appropriate.

%\vspace{-0.3cm}
\section{Outlook \& conclusion}
%\vspace{-0.3cm}

The data in Fig.\,\ref{fig_2} and \ref{fig_3} demonstrate that there are two coexisting energy scales in the HTSCs: one associated with the
superconducting $T_c$ and the other, as inferred primarily from the antinodal region properties, with the pseudogap $T^*$. The next most
critical step is that of addressing the subtle questions concerning the nature of these energy scales and the significance of the emerging
two-gap phenomenology towards the development of a microscopic description of high-$T_c$ superconductivity.

As for the pseudogap, which grows upon underdoping, it seems natural to seek a connection to the physics of
the insulating parent compound. Indeed, it has been pointed out that this higher energy scale might smoothly
evolve, upon underdoping, into the quasiparticle dispersion observed by ARPES in the undoped
antiferromagnetic insulator.\cite{filip,ronning} At zero doping the dispersion and quasiparticle weight in
the single-hole spectral function as seen by ARPES can be very well explained in terms of a self-consistent
Born approximation,\cite{sushkov} as well as in the diagrammatic quantum Monte Carlo \cite{Prokof'ev}
solution to the so called $t$-$t^{\prime}$-$t^{\prime\prime}$-$J$ model. In this model, as in the
experiment,\cite{filip,ronning} the energy difference between the top of the valence band at
($\pi/2$,$\pi/2$) and the antinodal region at ($\pi$,0) is a gap due to the quasiparticle dispersion of about
250$\pm$30\,meV. Note that this would be a single-particle gap $\Delta$. For the direct comparison with the
pseudogap data in Fig.\,\ref{fig_2}, we would have to consider $2\Delta\!\sim\!500$\,meV; this however is
much larger than the $x\!=\!0$ extrapolated pseudogap value of 186\,meV found from our analysis across the
phase diagram. Thus there  seem to be an important disconnection between the finite doping pseudogap and the
zero-doping quasiparticle dispersion.

The fact that the pseudogap measured  in ARPES and SIN experiments is only half the size of the gap in SIS,
STM, $B_{1g}$ Raman, and heat conductivity measurements, points to a pairing gap. So, although the origin of
the pseudogap at finite doping remains uncertain, we are of the opinion that it most likely reflects a
pairing energy of some sort. To this end, the trend in Fig.\,\ref{fig_2} brings additional support to the
picture discussed by many authors that the reduction in density of state at $T^*$ is associated with the
formation of electron pairs, well above the onset of phase coherence taking place at $T_c$ (see, e.g.
Ref.\,\onlinecite{kotliar} and \onlinecite{emery}). The pseudogap energy $E_{pg}\!=\!2\Delta_{pg}$ would then
be the energy needed to break up a preformed pair. To conclusively address this point, it would be important
to study very carefully the temperature-dependence of the ($\pi$,0) response below $T_c$; any further change
with the onset of superconductivity, i.e. an increase of $E_{pg}$, would confirm the two-particle pairing
picture, while a lack thereof would suggest a one-particle band structure effect as a more likely
interpretation of the pseudogap.

The lower energy scale connected to the superconducting $T_c$ (parabolic curve in
Fig.\,\ref{fig_2},\,\ref{fig_3}) has already been proposed by many authors to be associated with the
condensation energy,\cite{kotliar,kl,uemuraplot,emery} as well as with the magnetic resonance in
INS.\cite{zhang} One might think of it as the energy needed to take a pair of electrons out of the
condensate; however, for a condensate of charged bosons, a description in terms of a collective excitation,
such as a plasmon or roton, would be more appropriate.\cite{Uemura} The collective excitation energy would
then be related to the superfluid density and in turn to $T_c$. In this sense, this excitation would truly be
a two-particle process and should not be measurable by single-particle spectroscopies. Also, if the present
interpretation is correct, this excitation would probe predominantly the charge-response of the system;
however, there must be a coupling to the spin channel, so as to make this process neutron active (yet not as
intense as predicted by the sum rule for pure spin-$1/2$ magnetic excitations, which is consistent with the
small spectral weight observed by INS). As discussed, one aspect that needs to be addressed to validate these
conjectures is the surprising correspondence between $q\!=\!0$ and $q\!=\!(\pi$,$\pi$) excitations, as probed
by Raman and INS respectively.

We are thus led to the conclusion that the coexistence of two energy scales is essential for high-$T_c$
superconductivity, with the pseudogap defining the pairing strength and the other, always smaller than the
pseudogap, reflecting the superconducting condensation energy. This supports the proposals that the HTSCs can
not be considered as classical BCS superconductors, but rather are smoothly evolving from the BEC into the
BCS regime,\cite{35b,35c,35d} as carrier doping is increased from the underdoped to the overdoped side of the
phase diagram.

%\vspace{-0.3cm}
\section{Acknowledgments}
%\vspace{-0.3cm}

S.H. would like to thank the University of British Columbia for its hospitality. Helpful discussions with
W.N. Hardy, D.G. Hawthorn, N.J.C. Ingle, and K.M. Shen are gratefully acknowledged. This work was supported
by the Deutsche Forschungsgemeinschaft (SFB 277, TP B5), the Alfred P. Sloan Foundation (A.D.), CRC and CIFAR
Quantum Materials Programs (A.D. and G.A.S), CFI, NSERC, and BCSI. M.A.H. has been supported by the Advanced
Light Source (ALS) Doctoral Fellowship Program, Berkeley; ALS is supported by the Director, Office of
Science, Office of Basic Energy Sciences, of the U.S. Department of Energy under Contract No.
DE-AC02-05CH11231.

\bibliographystyle{unsrt}
\bibliography{Stefan_PG_RPP_04}

\begin{thebibliography}{10}
\bibitem{intrinsic}
{The most convincing tunneling results showing two coexisting gaps were
  actually obtained by intrinsic tunneling,\cite{17, 18, 19, 19a} in particular
  from Bi$_2$Sr$_2$CuO$_{6+\delta}$ (Bi2201).\cite{18} However, because this
  technique suffers from systematic problems,\cite{19a} and one would anyway
  have to scale the Bi2201 data because of the lower value of $T_c$ and in turn
  gap energy scales, these results were not included in Fig.\,\ref{fig_2}
  or\,\ref{fig_3}. Since intrinsic tunneling is in principle a clean SIS
  experiment which measures pair energies through Josephson tunneling, a
  refinement of the technique might provide an accurate estimate of both
  superconducting and pseudogap simultaneously, and is thus highly desirable.}
\end{thebibliography}

\end{document}